\documentclass[modern]{aastex62}

\usepackage{xspace}
\usepackage{}

\newcommand{\msun}{\mbox{$\mathrm{M}_{\sun}$}\xspace}
\newcommand{\mearth}{\mbox{$\mathrm{M}_{\oplus}$}\xspace}

\newcommand{\gaia}{\textit{Gaia}\xspace}

\newcommand\kepler{\emph{Kepler}\,}

\definecolor{linkcolor}{rgb}{0.1216,0.4667,0.7059}

\usepackage{xcolor, fontawesome}
\definecolor{twitterblue}{RGB}{64,153,255}
\newcommand\twitter[1]{\href{https://twitter.com/#1 }{\textcolor{twitterblue}{\faTwitter}\,\tt \textcolor{twitterblue}{@#1}}}

\graphicspath{{../notebooks/}}

\received{October 29, 2019}
\accepted{November 23, 2019}
\submitjournal{ApJ}

\shorttitle{No Massive Companion to GJ\,1151}
\shortauthors{B. J. S. Pope et al.}

\begin{document}

\title{No Massive Companion to the Coherent Radio-Emitting M~Dwarf GJ\,1151}

\correspondingauthor{Benjamin J. S. Pope \twitter{fringetracker}}
\email{benjamin.pope@nyu}

\author[0000-0003-2595-9114]{Benjamin J. S. Pope}
\affiliation{Center for Cosmology and Particle Physics, Department of Physics, New York University, 726 Broadway, New York, NY 10003, USA}
\affiliation{Center for Data Science, New York University, 60 Fifth Ave, New York, NY 10011, USA}
\affiliation{NASA Sagan Fellow}

\author[0000-0002-8171-8596]{Megan Bedell}
\affiliation{Center for Computational Astrophysics, Flatiron Institute, 162 Fifth Ave, New York, NY 10010, USA}

\author[0000-0002-7167-1819]{Joseph R. Callingham}
\affiliation{ASTRON, Netherlands Institute for Radio Astronomy, Oude Hoogeveensedijk 4, Dwingeloo, 7991 PD, The Netherlands}

\author[0000-0002-0872-181X]{Harish K. Vedantham}
\affiliation{ASTRON, Netherlands Institute for Radio Astronomy, Oude Hoogeveensedijk 4, Dwingeloo, 7991 PD, The Netherlands}
\affiliation{Kapteyn Astronomical Institute, University of Groningen, PO Box 800, NL-9700 AV Groningen, the Netherlands}

\author[0000-0003-1624-3667]{Ignas A. G. Snellen}
\affiliation{Leiden Observatory, Leiden University, Postbus 9513, 2300 RA Leiden, The Netherlands}

\author[0000-0003-0872-7098]{Adrian M. Price-Whelan}
\affiliation{Center for Computational Astrophysics, Flatiron Institute, 162 Fifth Ave, New York, NY 10010, USA}

\author[0000-0001-5648-9069]{Timothy W. Shimwell}
\affiliation{ASTRON, Netherlands Institute for Radio Astronomy, Oude Hoogeveensedijk 4, Dwingeloo, 7991 PD, The Netherlands}



\begin{abstract}
The recent detection of circularly polarized, long-duration ($>8$~hr) low-frequency ($\sim$150\,MHz) radio emission from the M4.5~dwarf GJ\,1151 has been interpreted as arising from a star-planet interaction via the electron cyclotron maser instability. The existence or parameters of the proposed planets have not been determined. Using  20 new HARPS-N observations, we put $99^{\text{th}}$-percentile upper limits on the mass of any close companion to GJ\,1151 at $M\sin{i} < 5.6~\mearth$. With no stellar, brown dwarf, or giant planet companion likely in a close orbit, our data are consistent with detected radio emission emerging from a magnetic interaction between a short-period terrestrial-mass planet and GJ\,1151. \href{https://github.com/benjaminpope/video}{\color{linkcolor}\faGithub} 

\end{abstract}


\section{Introduction} 
\label{sec:intro}

Exoplanet science has flourished over the last three decades. The number of known planets has doubled nearly every two years since 1995 \citep{Mamajek2016} and this accelerating rate of discovery is projected to continue for at least the next decade if current and upcoming space-based surveys deliver their expected results. However, despite extensive searches \citep[e.g.][]{2000ApJ...545.1058B,2013A&A...552A..65L,Lynch18,2015MNRAS.446.2560M} and the possible exception of a flare from $\epsilon$~Eri \citep{2018ApJ...857..133B}, neither exoplanets nor their host stars have been detected at radio frequencies, as the non-flaring emission of such systems is likely too faint for most current low-frequency telescopes with the exception of the  SKA-precursor LOFAR \citep[the LOw-Frequency ARray:][]{lofar}. LOFAR has unparalleled sensitivity at 150~MHz where these interactions are expected to emit significant radiation via the electron cyclotron maser instability \citep{2007P&SS...55..598Z,2018ApJ...854....7L}: with its orders-of-magnitude increase in sensitivity and survey speed, the detection of nearby stars and planets is a realistic prospect \citep{pope19}. 

High stellar UV flux and flaring are thought to pose serious problems for habitability of planets around M~dwarfs \citep{Shields16,2019AsBio..19...64T}, though this may not be sufficiently severe in comparison to the early Earth to prevent the emergence of life \citep{2019MNRAS.485.5598O}. The stellar wind potentially poses a more severe problem. Theoretical studies have disagreed on the extent to which a planetary magnetosphere provides protection for its atmosphere from stripping by the radiation or wind of the host star \citep[e.g.][]{2013ApJ...770...23Z,2016A&A...596A.111R,2017ApJ...844L..13G}. Star-planet magnetic interactions (SPMI) analogous to the electrodynamic interaction of Jupiter and Io are theorized to occur when the interaction with the magnetized stellar wind of the host star is sub-Alfv\'{e}nic (i.e. the Alfv\'{e}n wave speed is greater than the wind velocity). Under these conditions there is no bow shock separating the magnetospheres of the star and planet, and particles from the stellar wind reach much deeper into the planetary magnetosphere \citep{2014ApJ...790...57C}. This is thought to be the case for Proxima~Centauri~b \citep{2016ApJ...833L...4G} and the inner TRAPPIST-1 planets \citep{2017ApJ...843L..33G}. With stellar wind flux orders of magnitude higher than that received by Earth, this may be a leading-order effect for stripping otherwise-habitable exoplanets of their atmospheres. The energy flux to the stellar surface from such an SPMI may give rise to strong chromospheric lines at the magnetic connection footprint on the star \citep{cuntz2000,shkolnik2008,lanza13,luger17,2019arXiv190701020S,cauley19}, or to radio signals \citep{2007P&SS...55..598Z, 2013A&A...552A.119S}. Detections of radio emission from brown dwarfs \citep[e.g.][]{kao16,gagne17,kao18} have been interpreted as auroral, but have not so far been associated with exoplanet candidates. The search for the radio emission from M~dwarf planets is therefore a key component of understanding their long term evolution and habitability \citep{2017ApJ...849L..10B,2018ApJ...854...72T,2019arXiv190607089V}, but observational signatures of this have not so far been detected \citep[e.g.][]{Lynch18,Lenc18,pineda18}.

Rather than explicitly searching for radio emission from known exoplanet hosts, \citet{Callingham_2019} cross-matched sources identified by the LOFAR Two-meter Sky Survey \citep[LoTSS:][]{lotss} with nearby stellar sources found by \gaia, finding the great majority of matches to be chance associations with radio-bright active galactic nuclei. But by restricting the crossmatch to LoTSS sources to only those that display circularly-polarized emission, the rate of chance associations with background radio galaxies is dramatically reduced. Based on this restricted cross-match, \citet{vedantham20} report the detection of the quiescent M4.5 dwarf GJ\,1151 at low radio frequencies with LOFAR, with properties that suggest the low-frequency radio emission is driven by a star-planet magnetic interaction. While M~dwarfs are known to flare at low frequencies \citep[e.g.][]{Villadsen19}, this emission lasts over 8\,h and is $64\pm6\%$ circularly polarised. Such emission can be generated by the electron cyclotron maser instability (ECMI) through the interaction of the star with a short ($\sim 1-5$\,day) period planet or a close stellar companion as seen in interacting binaries such as UV Ceti or RS CVn systems \citep{2017ApJ...836L..30L}.

Since the radio emission implies a potential planetary or sub-stellar companion, but no such companion is previously known, in this Letter we present and analyze HARPS-N \citep[High Accuracy Radial velocity Planet Searcher:][]{harpsn} observations of GJ\,1151 in order to search for radial velocity signals of the proposed companions. We do not detect any planets, but place strong upper limits of a few Earth masses on the $M\sin{i}$ of any possible companions, ruling out any short-period massive objects or a close binary companion.  

\section{RV Data and Analysis}
\label{sec:rv}

We obtained 20 epochs of observations of GJ\,1151 with HARPS-N from 2018-12-20 to 2019-02-27, as a Director's Discretionary Time program (Program ID: A38DDT2; PI: Callingham). While RVs were extracted from these using the standard HARPS pipeline, its performance on this M4.5 dwarf was poor, resulting in a spurious RV scatter of several km/s. 

We therefore reprocessed these data using \textit{wobble} \citep{wobble}, a data-driven package which simultaneously non-parametrically constructs a stellar spectral template and telluric spectral components and uses these, rather than model spectral masks, to extract radial velocities. We used only the 30 reddest echelle orders, as the signal-to-noise in the blue orders was much too low to resolve spectral features in this extremely red star. We expect \textit{wobble} to perform favorably in this regime compared to the standard HARPS cross-correlation-based approach. This is because \textit{wobble} can extract RV information from telluric-contaminated spectral regions, which are increasingly prevalent in the red, and it needs no a priori knowledge of the underlying stellar spectrum, for which templates may be unreliable for such a late spectral type.

\begin{figure}
\plotone{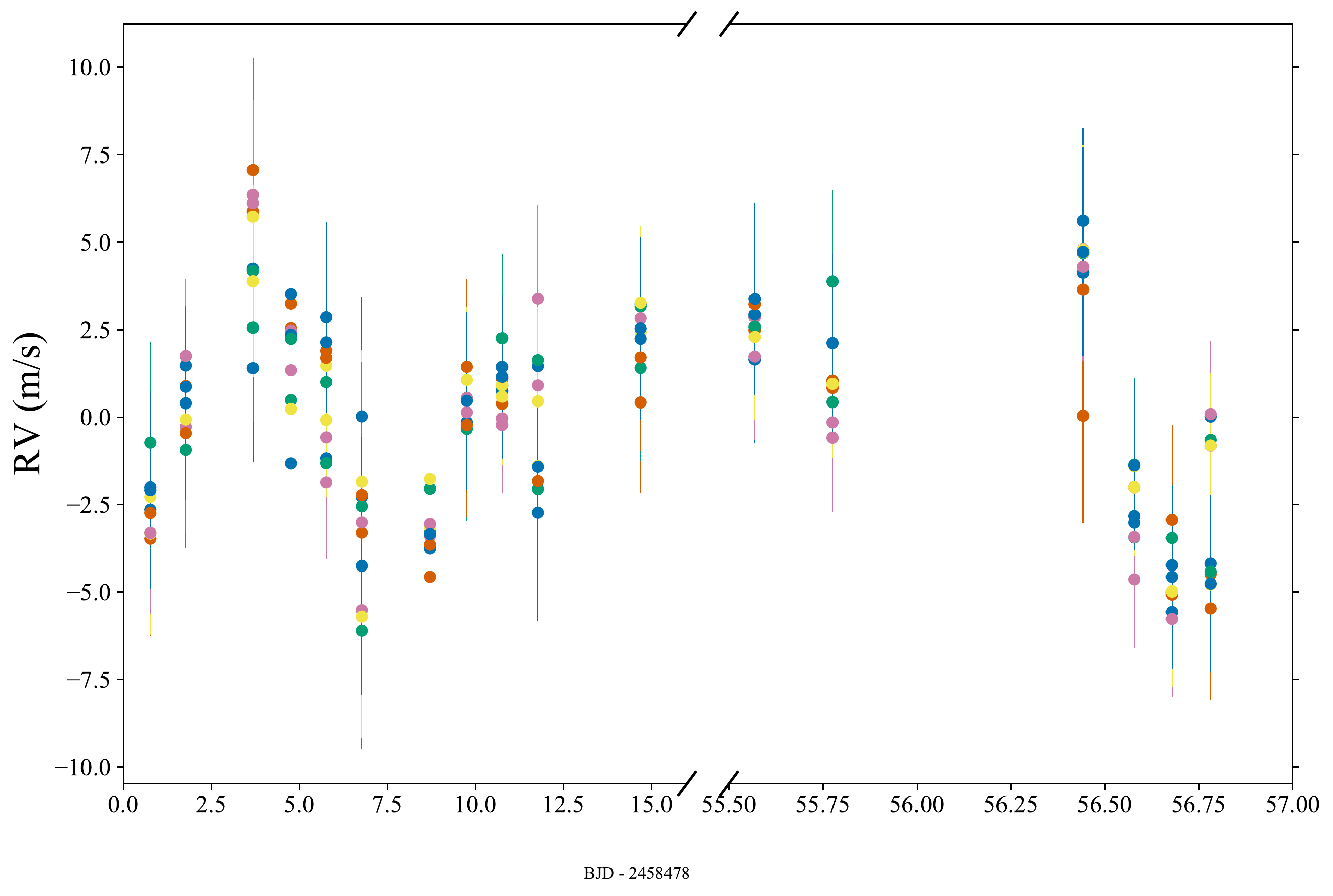}

\caption{\label{xvalidation}
Leave-one-out cross-validation of \textit{wobble} RVs. One epoch at a time is left out of the global model, and the results of processing the remaining epochs are shown in different colours. There is overall consistency between the different time series, with a diversity of order $\sim$ the quoted uncertainty between the individual realizations. 
}
\end{figure}

We found that the second epoch (2018-12-22) had a significantly higher extracted RV uncertainty than the others, and accordingly excluded this from the global \textit{wobble} model. In order to assess template-dependent systematic errors, we conducted a `leave-one-out' cross-validation, excluding one additional epoch at a time and rerunning \textit{wobble} to search for consistency between the outputs. As seen in Figure~\ref{xvalidation}, the different resulting time series are broadly consistent in their directions of deviation from the mean, with a scatter between them of order $\sim$ the quoted uncertainties. We therefore believe the uncertainties on the \textit{wobble} RVs are realistic but that they are also model-dependent systematics, and therefore likely correlated, though in subsequent analysis we will treat them as independent and Gaussian. 

We use \textit{The Joker} Bayesian RV analysis pipeline \citep{joker} with its default prior choices implemented in the new version 1.0 of the package \citep{pricewhelan19}. This pipeline is optimized for small numbers of irregularly-spaced observations, using importance sampling to fit Keplerian orbits to RV data and infer posterior planet parameters. 
We draw $10^{7}$ samples from a separable prior, such that all orbital elements are assumed to be independent. The prior over log-period is uniform over the range $\log P \in [0.5\,\text{d},8\,\text{d}]$; the prior over eccentricity is given by \citet{Kipping:2013}; the priors over velocity semi-amplitude $K$ and systemic velocity are assumed to be very broad and Gaussian (such that they are effectively uniform over the region that the likelihood has support); and all other priors are assumed to be uniform. We allow an additional astrophysical jitter (white noise) term to vary with a lognormal prior on jitter $\ln {(s/(\text{m/s}))^2} \sim {N}(1,2)$. We also include a term for a linear trend, which allows for very-long-period companions. 

\begin{figure}
\plotone{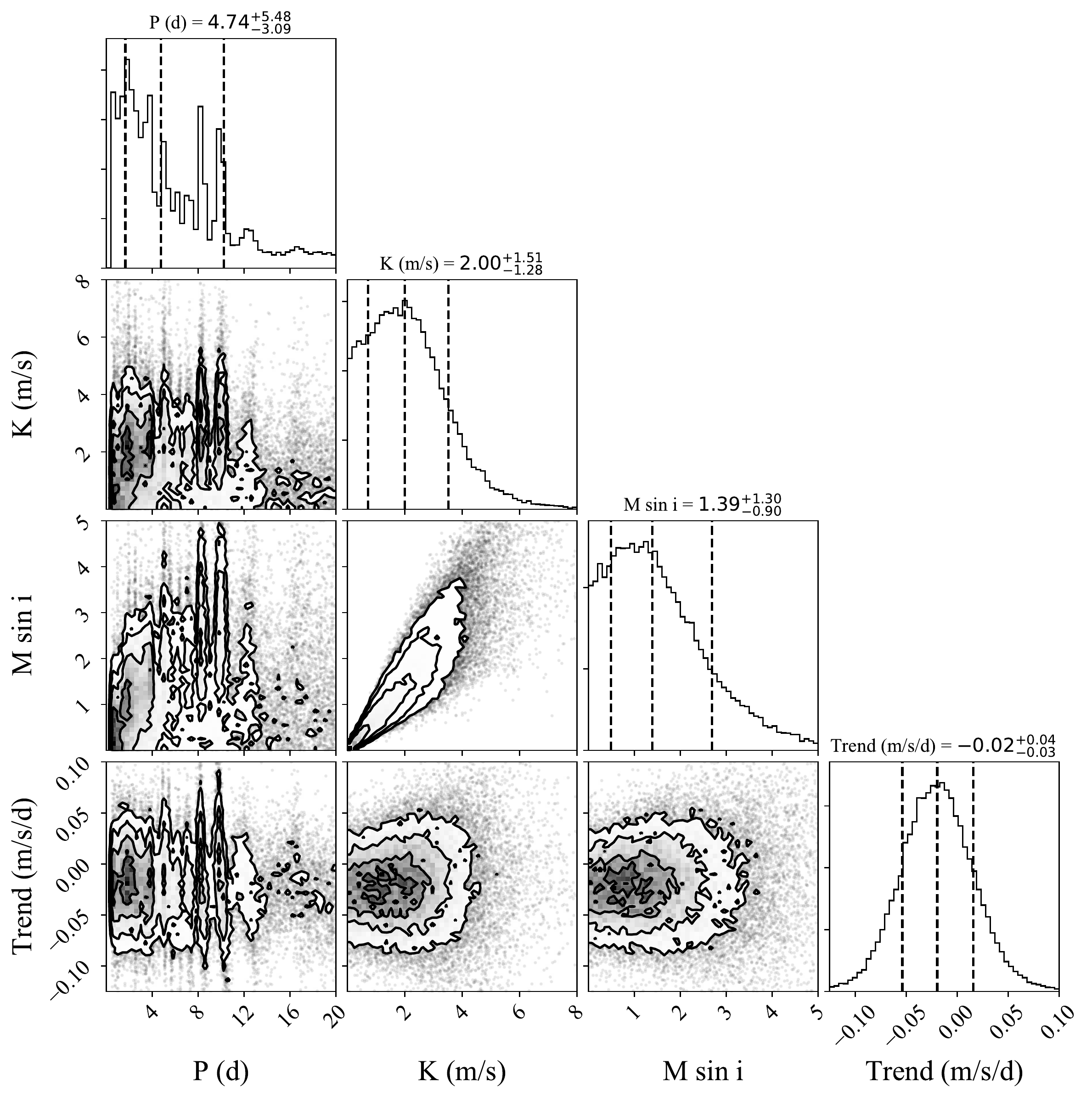}
\caption{\label{cornerplot}
Cornerplot of posterior samples from \textit{The Joker} for GJ\,1151, made using \texttt{corner.py} \citep{corner}. The RV trend, the RV semi-amplitude $K$, and $M\sin{i}$ of any companion, are all consistent with zero. The spread in the relation between $M\sin{i}$ and $K$ is due to the estimated uncertainties in the stellar mass.
}
\end{figure}

\begin{figure}
\plotone{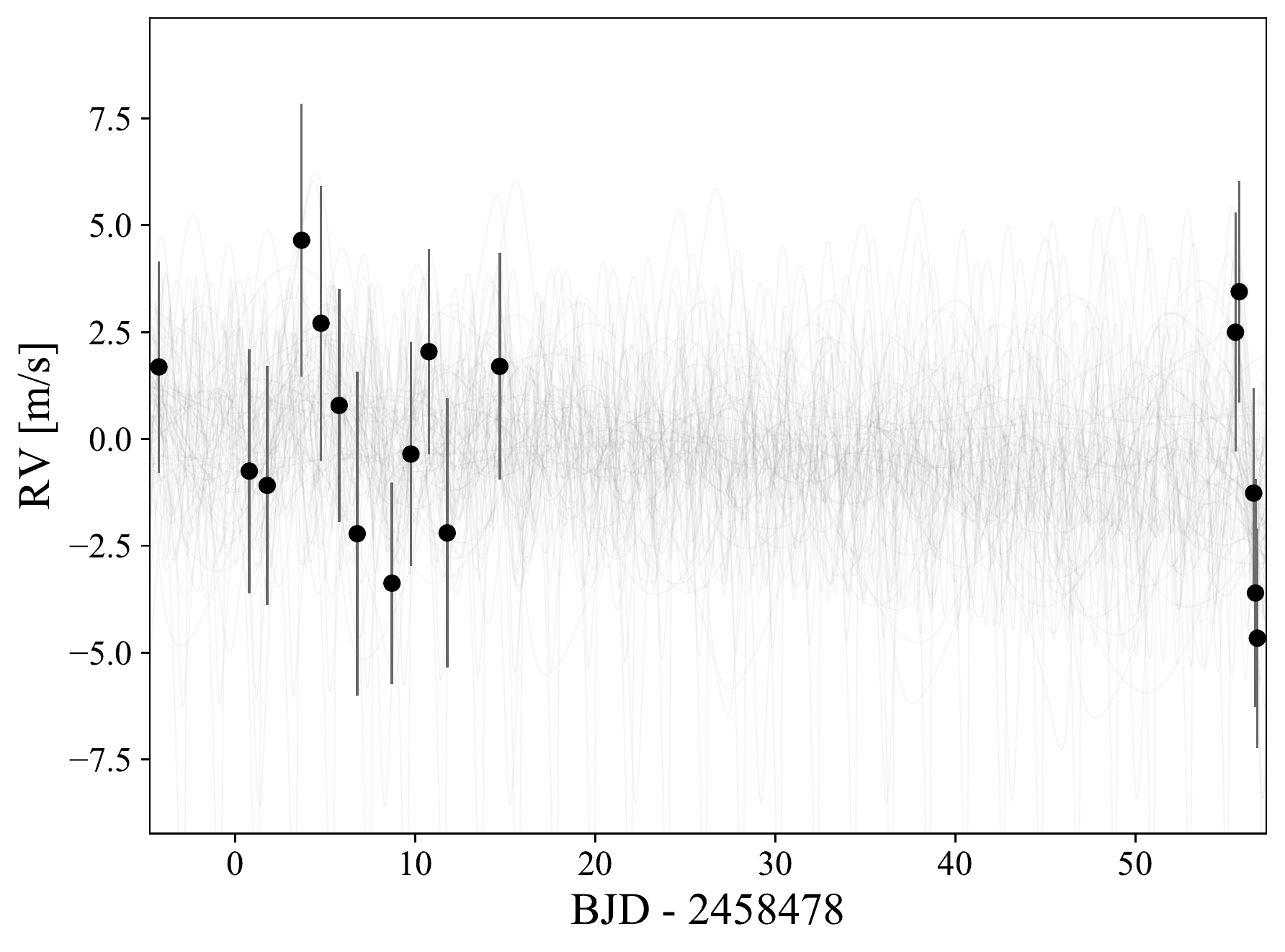}
\caption{\label{jokermodel1151}
RV time-series generated from\textit{The Joker} posterior samples overlaid on the HARPS-N data for GJ\,1151. We see no clear Keplerian fit.
}
\end{figure}

After rejection sampling, we obtain 32768 posterior samples, displayed in Figure~\ref{cornerplot}. In Figure~\ref{jokermodel1151} we show posterior model draws overlaid on the data: it is clear that no good fit is found. We place $99^{th}$-percentile upper limits on the RV semi-amplitude $K < 7.7~\text{m}/\text{s}$, which translates to $M\sin{i} < 5.6 \mearth{}$, using a stellar mass of $0.167 \pm 0.025\,\msun{}$ \citep[as determined by][and estimated Gaussian uncertainties]{2016yCat..18210093N}. The posterior for the RV trend of $-0.02^{+0.04}_{-0.03}$ $\text{m}/\text{s}\,\text{d}^{-1}$ is consistent with zero, providing no evidence for a long-period massive companion. The source code and data for our calculations are available online at \href{https://github.com/benjaminpope/video}{github.com/benjaminpope/video}.

\section{Discussion and Outlook}
\label{sec:discussion}

As part of the CARMENES project \citep{carmenes}, high angular resolution observations of of GJ~1151 have been obtained lucky imaging instrument FastCam \citep{2017A&A...597A..47C}, ruling out a stellar companion at separations greater than $\sim$1~au. These new RV data fill in this gap to short periods, indicating no massive companion except if it is in a face-on orbit, which is unlikely a priori. 

Planets are common around such M~stars such as GJ~1151: using \kepler, \citet{2019arXiv190505900H} estimate that M3-M5 dwarfs host $1.19^{+0.70}_{−0.49}$ planets per star with radii from $0.5-2.5 R_\oplus$ and periods from  0.5-10\,d.  The results of this analysis indicate no binary companions or planets on short orbits more massive than Neptune as the origin of the radio signal from GJ\,1151. We nevertheless cannot rule out in either case planets less massive than a few Earth masses, or highly-inclined short period orbits. The non-detection of a planet is therefore less important than the exclusion of \emph{non}-planetary models, such as emission from a stellar binary interaction. 


\citet{vedantham20} derive an approximate mass and period estimate for their planet candidate based on the energetics of the SPMI. In their benchmark model, an Earth-like planet in a $\sim 1-5$\,day orbit can generate the observed emission within a reasonable range of interaction and emission efficiencies.
A planet with a larger radius $r$ has a larger cross-section for wind interaction $\propto r^2$, while the stellar Poynting flux at the location of the planet drops with semi-major axis\footnote{Assuming a Parker-spiral configuration for the stellar magnetic field.} $a$ as $\approx a^{-2}$, so that the radio flux provides a lower limit on $r/d$. Given a planetary mass scaling $\propto r^3$ and orbital period $\propto a^{\frac{2}{3}}$, the radio detection provides a lower limit on $m/p^2$. Hence, at sufficiently high efficiencies, even a sub-Earth-mass planet is plausible.

While the data presented in this Letter conclusively rule out stellar and gas-giant companions, there is a substantial region of parameter space for terrestrial planetary companions that cannot be excluded and the star-planet interaction hypothesis remains reasonable. Furthermore, from the radio observations of \citet{vedantham20} alone, it is possible that the emission from the GJ~1151 system could originate directly from an exoplanet's magnetosphere. However, such an emission site would imply an unreasonably strong magnetic field for a terrestrial-sized planet, with only hot~Jupiter magnetic fields considered to be on the order of tens of gauss \citep{cauley19}. For comparison, M~dwarfs are known to possess magnetic fields on the order of tens of gauss and greater \citep{2008MNRAS.390..567M}. Therefore, the present work disfavors radio emission directly from an exoplanet magnetosphere unless a terrestrial planet can generate a much stronger magnetic field than has previously been considered. We propose it is likely that the emission is from a star-planet interaction.

Because GJ~1151 is so red, to achieve significantly higher precision than attained by HARPS would require moving to the infrared, using an IR precision RV instrument such as the Habitable-zone Planet Finder \citep[HPF:][]{hpf}, SPIRou \citep{spirou}, or CARMENES \citep{carmenes}, by which GJ\,1151 is already subject to monitoring \citep{carmenesinput}.

\section*{Acknowledgements} 

This work was performed in part under contract with the Jet Propulsion Laboratory (JPL) funded by NASA through the Sagan Fellowship Program executed by the NASA Exoplanet Science Institute. This article is based on observations made in the Observatorios de Canarias del IAC with the Telescopio Nazionale Galileo (TNG) operated on the island of La Palma by the Fundaci\'{o}n Galileo Galilei of the Istituto Nazionale di Astrofisica (INAF) at the Spanish Observatorio del Roque de los Muchachos of the Instituto de Astrof\'{i}sica de Canarias. We thank the director of the TNG for awarding this program (Program ID: A38DDT2) Director's Discretionary Time. I.S. acknowledges funding from the European Research Council (ERC) under the European Union's Horizon 2020 research and innovation program under grant agreement No 694513.

BJSP acknowledges being on the traditional territory of the Lenape Nations and recognizes that Manhattan continues to be the home to many Algonkian peoples. We give blessings and thanks to the Lenape people and Lenape Nations in recognition that we are carrying out this work on their indigenous homelands.

This research made use of NASA's Astrophysics Data System and the SIMBAD database, operated at CDS, Strasbourg, France. Some of the data presented in this paper were obtained from the Mikulski Archive for Space Telescopes (MAST). STScI is operated by the Association of Universities for Research in Astronomy, Inc., under NASA contract NAS5-26555. Support for MAST for non-HST data is provided by the NASA Office of Space Science via grant NNX13AC07G and by other grants and contracts. 

\software{\textit{wobble} \citep{wobble};\,\textit{The Joker} \citep{joker};\, \texttt{corner.py} \citep{corner};\,\textsc{IPython} \citep{PER-GRA:2007};\, SciPy \citep{scipy};\, and Astropy, a community-developed core Python package for Astronomy \citep{astropy}.}



\bibliography{ms}



\end{document}